\documentclass[sigconf]{acmart}
\AtBeginDocument{%
  }

\usepackage{booktabs} 
\usepackage{tabularx} 

\setcopyright{none}
\acmYear{}
\copyrightyear{}
\acmConference{}{}{}
\acmBooktitle{}
\acmDOI{}
\acmISBN{}

\begin{document}

\title{Quantum State Discrimination Enhanced by FPGA-Based AI Engine Technology}


\author{Anastasiia Butko}
\affiliation{%
  \institution{Lawrence Berkeley National Laboratory}
  \city{Berkeley, CA}
  \country{USA}}
\email{AButko@lbl.gov}
\thanks{Corresponding author: Anastasiia Butko (abutko@lbl.gov).}

\author{Artem Marisov}
\affiliation{%
  \institution{System View Inc}
  \city{San Mateo, CA}
  \country{USA}}
\email{marisov.av@gmail.com}

\author{David I. Santiago}
\affiliation{%
  \institution{Lawrence Berkeley National Laboratory}
  \city{Berkeley, CA}
  \country{USA}}
\email{David.I.Santiago@lbl.gov}

\author{Irfan Siddiqi}
\affiliation{%
  \institution{Lawrence Berkeley National Laboratory}
  \city{Berkeley, CA}
  \country{USA}}
\email{IASiddiqi@lbl.gov}



\begin{abstract}
  Identifying the state of a quantum bit (qubit), known as quantum state discrimination, is a crucial operation in quantum computing. However, it has been the most error-prone and time-consuming operation on superconducting quantum processors. Due to stringent timing constraints and algorithmic complexity, most qubit state discrimination methods are executed off-line. In this work, we present an enhanced real-time quantum state discrimination system leveraging FPGA-based AI Engine technology. A multi-layer neural network has been developed and implemented on the AMD Xilinx VCK190 FPGA platform, enabling accurate in-situ state discrimination and supporting mid-circuit measurement experiments for multiple qubits. Our approach leverages recent advancements in architecture research and design, utilizing specialized AI/ML accelerators to optimize quantum experiments and reducing the use of FPGA resources.
\end{abstract}




\maketitle

\section{Introduction}

Quantum computing has gained significant attention over the past decades due to its remarkable potential to solve complex problems that are difficult for classical computers. Applications of quantum computing address a variety of fields, including cryptography, optimization, drug discovery, and materials science~\cite{preskill2018quantum}.

Various quantum technologies are making progress towards practical applicability, with superconducting qubits leading in terms of coherence times, scalability, and integration with established fabrication techniques~\cite{bourassa2020high}. Recent advancements in superconducting qubit technology are highlighted by significant milestones such as Google's demonstration of quantum supremacy~\cite{arute2019quantum} and IBM's progress in increasing quantum volume~\cite{ibm2020quantum}, along with ongoing innovations in error-correcting codes~\cite{bausch2024learning}. However, despite these achievements, considerable challenges remain that must be overcome to fully harness the potential of quantum computing.

One of the challenges lies in the quantum control architectures that predominantly rely on classical hardware. Recent literature identifies numerous issues inherent to quantum systems, including the need for simultaneous control of individual qubits with unique properties, stringent latency and clock speed requirements, the transient nature of qubit coherence, and the complexities associated with error correction and decoding~\cite{2019arXiv191205114R}. These challenges require the development of novel, quantum-specific solutions that have not been extensively explored within the classical computing community.

An essential function in quantum computing is mid-circuit measurement (MCM)~\cite{corcoles2019generating}, a technique that enables the measurement of qubit states at various points throughout a quantum computation. This capability is critical for implementing adaptive quantum algorithms and enhancing error mitigation strategies, allowing systems to dynamically respond to measurement outcomes. Quantum state discrimination is integral to the process of MCM, as it involves accurately determining the state of qubits with high speed and efficiency, thus improving the overall fidelity of computations.

Recent works have made significant advances in the implementation of MCM in superconducting qubits. Some studies focus on developing custom register-transfer level (RTL) designs for execution on FPGA logic, while others investigate acceleration technologies designed to address the unique challenges of quantum state discrimination. These technologies can deliver favorable trade-offs in terms of speed, accuracy, and resource utilization, alongside the potential for scalability to larger quantum systems.

In this study, we concentrate on improving quantum state discrimination through the use of recent specialized accelerator technologies, such as the AMD AI Engine~\cite{amd2020ai}. The integration of FPGA-based AI engines within the quantum computing framework represents a promising direction for overcoming the control complexities associated with quantum system scaling and resource utilization.

\begin{figure*}[htbp]
  \centering
  \includegraphics[width=\linewidth, trim={15 0 15 0},clip]{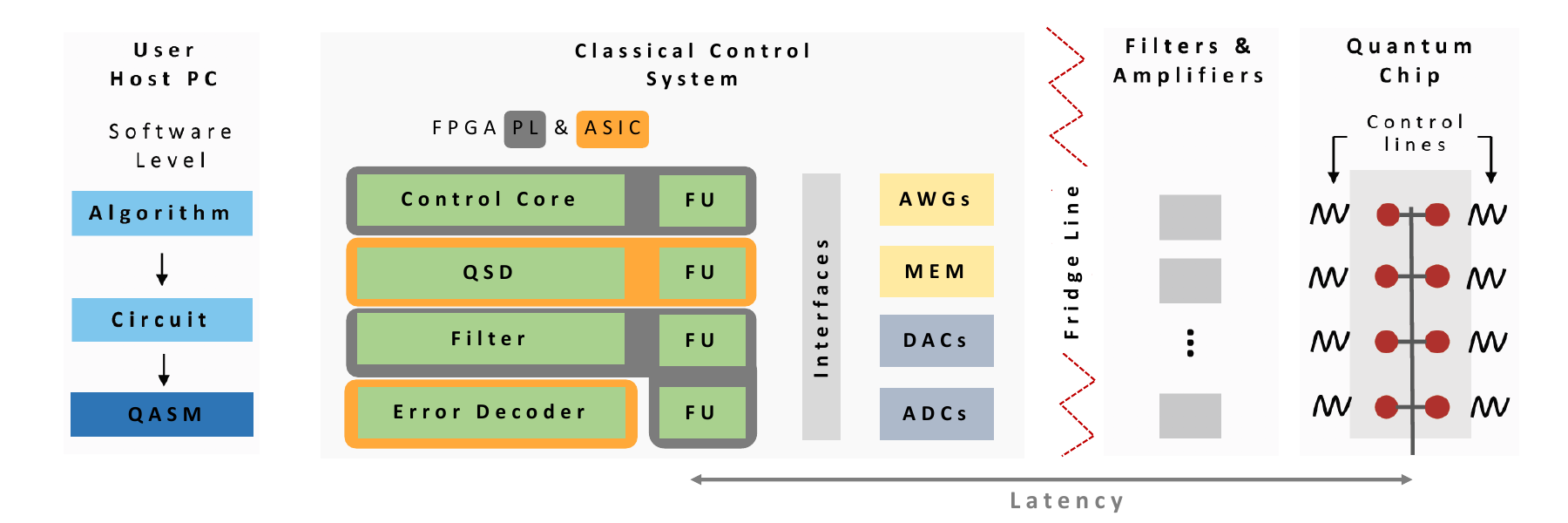}
  \caption{Architecture of the Classical Control System: On the left, the User Host PC depicts the software-level interactions, starting from the high-level algorithm through to circuit representation and Quantum Assembly Language (QASM).
  The central section shows the architecture of the classical control system, comprising a Control Core and functional units (FUs), Quantum State Discriminator (QSD), filtering, and error decoding. Interfaces facilitate communication with various components.
  On the right, the diagram highlights the downstream components such as Arbitrary Waveform Generators (AWGs), memory modules (MEM), Digital-to-Analog Converters (DACs), and Analog-to-Digital Converters (ADCs) which interact with the Filters \& Amplifiers before connecting to the quantum chip.}
  \label{fig:arch}
\end{figure*}

This paper makes the following technical contributions:

\begin{itemize}
    \item Novel Implementation: This work introduces a novel implementation of real-time quantum state discrimination utilizing the Versal AI platform.
    \item Neural Network Architecture Mapping: The paper details the mapping of a neural network architecture onto the vector units of the AI Engine.
    \item Practical Demonstration: A practical demonstration of the proposed approach is provided, showcasing its feasibility and effectiveness.
    \item Efficiency and Resource Utilization: Results indicate that the processing time is competitive with state-of-the-art solutions, while utilizing only a small percentage of the available resources, thus allowing for significant optimization and scalability potential.
\end{itemize}

The rest of the paper is organized as follows. Section \ref{sec:background} provides background on real-time control architecture and the role of quantum state discrimination. Section \ref{sec:related} describes the related work. In Section \ref{sec:impl}, we describe the proposed implementation details. Section \ref{sec:results} reports evaluation methodology and comparison results. Finally, Section \ref{sec:conc} concludes this paper and describes future work. 
 
\section{Background}
\label{sec:background}

Control hardware for superconducting qubits bridge the gap between classical instructions and quantum processors by generating precise microwave pulses and performing high-speed readout. FPGAs (Field-Programmable Gate Arrays) are the standard for these systems due to their ability to process signals in parallel with sub-microsecond latency. 

Commercial control stacks provide integrated "lab-in-a-box`` solutions that prioritize reliability, high channel density, and user-friendly software abstractions. Open-source platforms provide cost-effective and highly customizable alternatives for research labs~\cite{shammah2024open}.



\begin{table*}[htbp]
    \centering
    \caption{Trade-offs of Quantum State Discrimination Techniques}
    \label{tab:discrimination_tradeoffs}
    \begin{tabularx}{\textwidth}{l >{\raggedright\arraybackslash}X >{\raggedright\arraybackslash}X}
        \toprule
        \textbf{Technique} & \textbf{Advantages (Pros)} & \textbf{Challenges (Cons)} \\
        \midrule
        \textbf{Helstrom Bound} & Establishes the fundamental theoretical limit for minimum-error discrimination between two states. & Serves as a mathematical benchmark rather than a direct implementation for experimental signals. \\
        \addlinespace
        \textbf{LDA (Linear)} & Minimal computational latency and low resource overhead; highly effective for well-separated IQ clusters. & Performance degrades significantly when state distributions exhibit unequal covariance or high overlap. \\
        \addlinespace
        \textbf{QDA (Quadratic)} & Accommodates asymmetric noise distributions by allowing non-linear decision boundaries. & Increased parameter estimation requirements; higher classical computational cost compared to linear methods. \\
        \addlinespace
        \textbf{Machine Learning} & Capable of capturing complex, non-linear features in raw signal data to maximize readout fidelity. & Demands significant training data; inference must be optimized for real-time integration into feedback loops. \\
        \bottomrule
    \end{tabularx}
\end{table*}

\subsection{Real-time Control Architecture}

Figure~\ref{fig:arch} illustrates a high-level block diagram of a classical control system architecture that is based on FPGAs. The central section highlights various control components operating at both the digital and analog levels. Typically, the FPGA system includes three primary types of resources: Processing System (PS), Programmable Logic (PL), and additional accelerators, which may vary depending on the board configuration.

The PS typically consists of a microprocessor or embedded CPU that manages high-level tasks and system orchestration. It is responsible for executing control algorithms, communication with external systems, and data processing tasks that do not require real-time execution. The PS serves as the system's main control unit, facilitating seamless integration with software components shown on the left of Figure~\ref{fig:arch}.

The PL is comprised of reconfigurable hardware resources that allow for the implementation of specific functions, such as signal processing and custom control algorithms. The PL can be configured and adapted to meet the specific requirements of a quantum experiment, including real-time signal acquisition and control tasks.

Depending on the board configuration, additional accelerators (such as dedicated AI engines or DSPs) can be incorporated into the system to enhance performance. These accelerators are optimized for particular tasks, such as machine learning inference or complex numerical computations, thus significantly improving overall system efficiency and reducing processing latencies.

In the context of a quantum experiment, resources available on a single FPGA must be utilized and distributed judiciously. It is essential to identify which tasks are best suited for the PL and which can be offloaded to additional accelerators. Even when controlling a chip with as few as 8 to 16 qubits, the demands on PL logic can be significant, potentially straining the system's ability to meet timing requirements at high clock frequencies.

In this paper, we propose offloading the quantum state discrimination task to the AI Engine rather than using a functional block that consumes valuable PL resources~\cite{2024arXiv240618807V}. If successful, this approach will (i) free up PL resources, (ii) enhance scalability for additional tasks, and (iii) improve accuracy, as an FPGA PL implementation often necessitates thorough optimization, which can sometimes compromise the quality of state discrimination.

However, potential challenges to address include the efficient streaming of data acquired from qubits to the AI Engine, as well as ensuring that the responses are communicated back to the PL components to continue the mid-circuit measurement algorithm. It presents a formidable engineering challenge, as quantum operations, including readout, occur on timescales as short as 15–100 ns~\cite{PhysRevApplied.10.034040}.

\subsection{Quantum State Discrimination}

\begin{figure}[htbp]
  \centering
  \includegraphics[width=\linewidth, trim={20 0 20 0},clip]{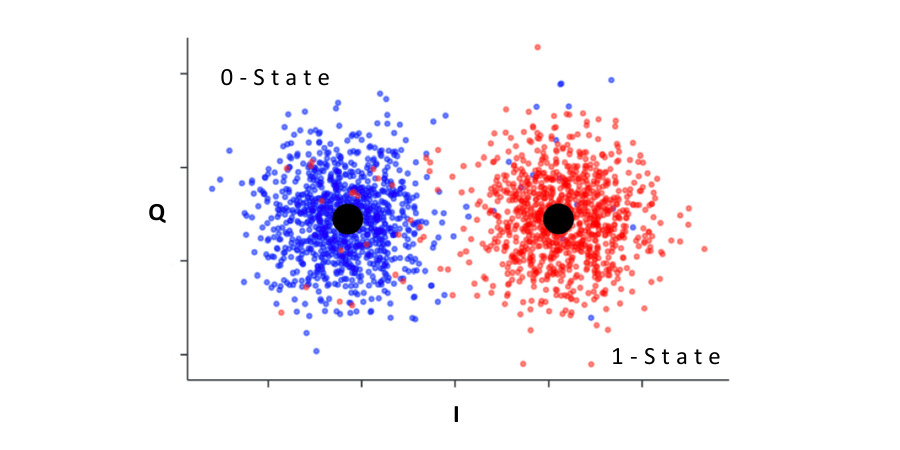}
  \caption{Visualization of Qubit State Readout in the IQ Plane. The figure displays clusters of data points representing the In-phase (I) and Quadrature (Q) components of the readout signal during the measurement of a qubit's state. The blue points correspond to the 0-State, while the red points correspond to the 1-State. Each cluster is centered around a mean value, indicated by the black dots, illustrating the distinct distributions of measurement outcomes for the two quantum states.}
  \label{fig:qsd}
\end{figure}

When a qubit is measured using dispersive readout, the resulting signal is a complex voltage that can be plotted on a 2D graph (the IQ plane). Figure~\ref{fig:qsd} illustrates qubit state readout in the IQ plane~\cite{2024arXiv240209532C}.

\textbf{Higher-Dimensional Systems.} The shift from binary (qubit) to multi-valued quantum systems is a primary area of focus for improving the efficiency and stability of quantum processors~\cite{goss2024extending}.  \textit{Qutrit (Quantum Trit)} is a unit of quantum information with three possible states ($|0\rangle, |1\rangle,$ and $|2\rangle$) that can exist in superposition. \textit{Qudit (Quantum Dit)} is a generalized $d$-dimensional quantum system where $d > 2$. Thus, when measuring a qutrit, the IQ plane visualization features three clusters rather than two.

Implementing efficient state discrimination for two-level systems through neural networks and custom ASICs will provide the high-speed, scalable classification framework necessary to resolve the increasingly complex, overlapping signal clusters of higher-dimensional qutrits and qudits in future quantum architectures.

\begin{figure*}[h]
  \centering
  \includegraphics[width=\linewidth, trim={10 30 10 30},clip]{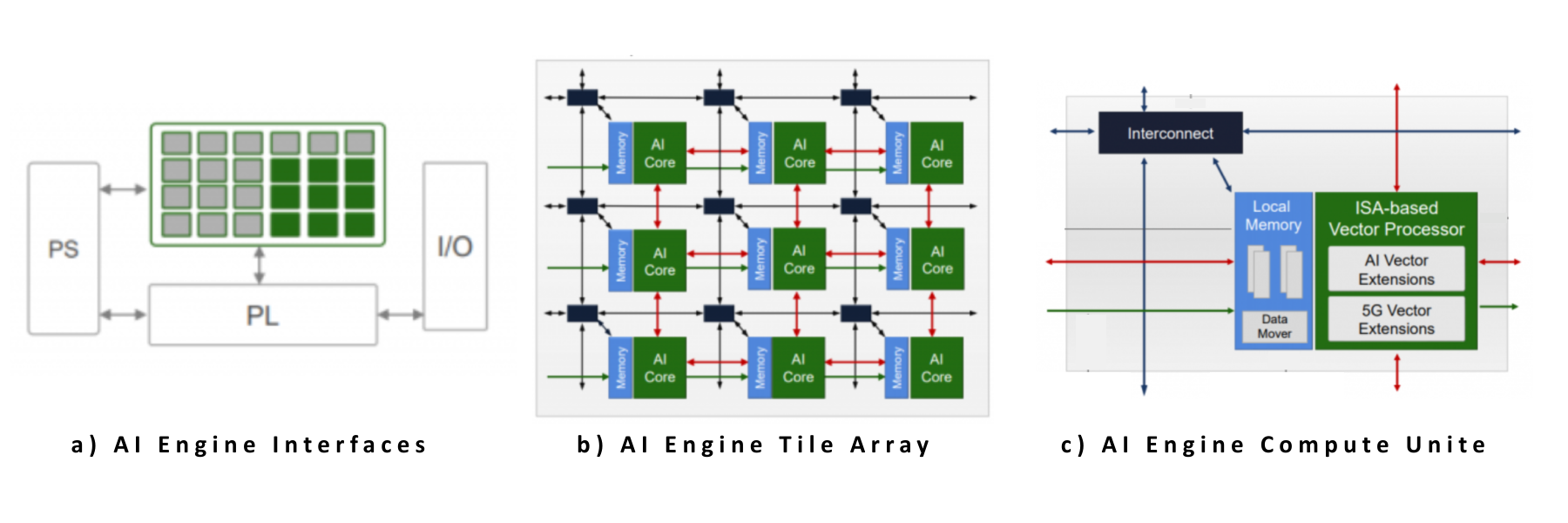}
  \caption{Overview of the AI Engine Architecture: (a) AI Engine Interfaces illustrating connections between the Processing System (PS), Programmable Logic (PL), and Input/Output (I/O); (b) Configuration of AI Engine Tiles featuring interconnected AI cores and memory resources; (c) Detailed structure of the AI Engine Core, showing the local memory, ISA-based vector processor, and interconnects for data movement and vector processing capabilities~\cite{xilinx2023versal}.}
  \label{ref:aie}
\end{figure*}

\textbf{Discrimination Techniques.}
To assign a specific measurement (a point in the IQ plane) to a state, researchers use various mathematical "lines`` or boundaries to separate the clusters.

Choosing a quantum state discrimination technique involves a trade-off between accuracy (fidelity), speed (latency), and computational complexity. The ideal method depends heavily on the specific hardware, noise levels, and the number of states being classified.

\begin{itemize}
    \item \textbf{Helstrom Bound}: This is the fundamental theoretical limit on how well two non-orthogonal states (represented by these clusters) can be distinguished~\cite{helstrom1976quantum}.
    \item \textbf{Linear Discriminant Analysis (LDA)}: Draws a straight line between two clusters to maximize separation~\cite{magesan2015machine}.
    \item \textbf{Quadratic Discriminant Analysis (QDA)}: Uses a curved boundary, which is more effective if the clusters have different shapes or spreads~\cite{ghojogh2019linear}.
    \item \textbf{Machine Learning (ML)}: Techniques like Support Vector Machines (SVM) or Neural Networks are used for high-fidelity discrimination when blobs are complex or overlapping~\cite{lienhard2022deep}.
\end{itemize}

Table~\ref{tab:discrimination_tradeoffs} summarizes the advantages and challenges of quantum state discrimination techniques. Over the past years, ML techniques have moved from theoretical explorations to essential tools for high-fidelity state discrimination.

\section{Related Work}
\label{sec:related}
Recent works in ML state discrimination focus on overcoming hardware noise, reducing readout latency through edge computing, and handling multi-qubit crosstalk. Recent research has prioritized moving ML models closer to the quantum hardware to achieve the sub-microsecond latencies required for quantum error correction (QEC).

Lienhard et al. demonstrated that deep neural networks (DNNs) significantly outperform traditional boxcar filters and linear discriminators in five-qubit systems~\cite{2022PhRvP..17a4024L}. Baumer et al. introduced a dynamic decoupling strategy for mid-circuit measurements. However, their method necessitated longer readout durations of 1.2 us and a feed-forward time of 650 ns~\cite{PRXQuantum.5.030339}. The QubiCML framework~\cite{2024arXiv240618807V} demonstrates the efficacy of hardware-accelerated discrimination by deploying a multilayer neural network on an RFSoC FPGA (Xilinx ZCU216). This architecture achieves a 98.5\% classification accuracy with a minimal inference latency of 54 ns, facilitating the low-latency feedback required for real-time mid-circuit measurements.

Building upon the successes of QubiCML, this work explores the implementation of high-fidelity state discrimination using AIE technology. We posit that as quantum processors scale, the inherent resource constraints of traditional PL will necessitate a transition toward heterogeneous architectures. By offloading compute-intensive classification tasks to the AIE, we aim to provide a more scalable trade-off between resource utilization and processing throughput, directly addressing the hardware bottlenecks of next-generation quantum control systems.


\section{Implementation}
\label{sec:impl}

AI engines are designed to handle the high-throughput demands of AI applications, particularly when it comes to vector operations, which are fundamental to various algorithms in neural networks, data processing, and mathematical computations.

Implementing algorithms on AIE involves navigating a range of complexities. Developers must consider effective data management, resource allocation, and memory access patterns to ensure optimal performance. Proper alignment of data structures can enhance throughput, while efficient utilization of on-chip resources minimizes latency and bottlenecks.

\begin{figure*}[h]
  \centering
  \includegraphics[width=\linewidth, trim={0 6cm 0 6cm},clip]{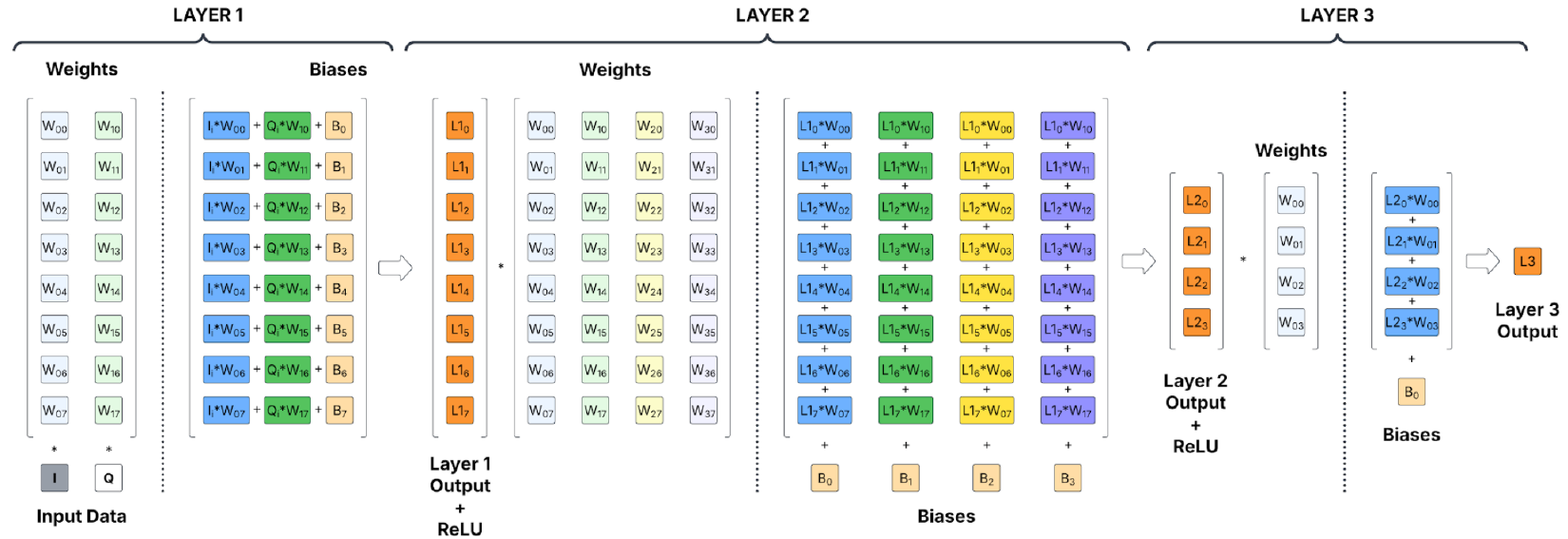}
  \caption{Architecture of a Multi-Layer Neural Network. In Layer 1, the input data is represented as a combination of In-phase (I) and Quadrature (Q) components. The data is processed through weighted connections and biases, resulting in the Layer 1 output after applying the ReLU activation function. Layer 2 takes the output from Layer 1 and again processes it using its own set of weights and biases, culminating in the Layer 2 output. Finally, Layer 3 further processes this output to yield the final output $L_3$, incorporating its own weights and biases. Each layer demonstrates how neural networks compute outputs through weighted summations and activation functions, illustrating the fundamental principles of neural network architecture.}
  \label{fig:nna}
\end{figure*}

\subsection{AI Engine Architecture}

Implementing quantum state discrimination on AIE represents a significant shift from traditional FPGA-based programming to a massively parallel, VLIW (Very Long Instruction Word) architecture. Unlike the fine-grained, bit-level control of PL, the AIE is a tiled array of high-performance RISC processors optimized for digital signal processing (DSP) and fixed-point/floating-point vector operations.

Figure~\ref{ref:aie} shows the AIE architecture. The AI Engine interfaces (Figure 1a) establish connectivity between multiple system components such as the PS, PL, and Input/Output (I/O) modules. On the VCK190 evaluation board~\cite{xilinx2023vck190}, the AIE array interacts with other system domains through a complex network of internal interfaces. The Network on Chip (NoC) is the primary backbone for data flow. It manages communication between the AIE array and external memory (DDR4/LPDDR4) or the high-performance PS. The PL-AIE communication is supported through high-speed streaming interfaces and memory-mapped AXI connections that allow the PL to act as a "data mover``. It can feed raw data from board peripherals (like HDMI or SFP28) directly into the AIE for acceleration. Finally, the VCK190 supports a maximum of 32 GMIO inputs and 32 GMIO outputs. These channels provide a direct link from the AIE to the NoC, enabling high-bandwidth access to system memory without taxing the PL fabric.

The AIE tile array shown in Figure~\ref{ref:aie} b) is a 2D mesh of 400 high-performance vector processors. These tiles are organized in a grid—typically 8 rows by 50 columns. Each of the 400 tiles is a self-contained compute unit consisting of three primary modules, i.e. AIE core, memory module, and interconnect module (Figure~\ref{ref:aie} c)). The core is a 7-way VLIW and Single Instruction Multiple Data (SIMD) processor. It includes a scalar unit and a powerful vector unit capable of executing fixed-point and floating-point operations.

For data storage and movement, every tile contains 32 KB of local data memory divided into eight banks, though a key architectural advantage is that each core can directly access the memory of its immediate neighbors in all four cardinal directions, creating a shared local memory pool. Beyond this local sharing, a sophisticated interconnect module handles global data movement via AXI4-Stream switches that support both circuit-switched and packet-switched routing. Additionally, dedicated cascade streams allow tiles to pass accumulator results directly to adjacent tiles, which is particularly efficient for chaining complex filters or matrix operations without taxing the primary streaming fabric.

\subsection{Neural Network Architecture}

Figure \ref{fig:nna} illustrates the architecture of a multi-layer neural network based on the FPGA implementation in~\cite{2024arXiv240618807V}. Each layer includes a set of weights and biases that are essential for transforming input data as it passes through the network.

\textbf{Layer 1} accepts the input data, denoted as \(Q\), and applies a linear transformation using its corresponding weights (\(W_{ij}\)) and biases (\(B_i\)). The output of this layer, indicated as \(L_1\), is computed by performing a matrix operation where the input data is multiplied by the respective weights and then added to the biases. This output is subsequently passed through a Rectified Linear Unit (ReLU) activation function, introducing non-linearity to the model.

\textbf{Layer 2} operates similarly, taking the output from Layer 1 as its input. The process involves multiplying this layer's weights (\(W_{ij}\)) with the Layer 1 output and adding the layer-specific biases (\(B_j\)). The result, represented as \(L_2\), is once again subjected to the ReLU activation function, enabling the network to capture more complex patterns in the data.

\textbf{Layer 3}, the final layer, processes the output from Layer 2 in the same manner. The transformation involves the multiplication of the weights associated with Layer 3 and the addition of biases to yield the final output, \(L_3\). At this stage, the network can provide predictions or classifications based on the learned features, effectively utilizing the transformations applied through each preceding layer.

Overall, the multi-layer architecture allows for progressively deeper feature extraction, with each layer contributing to an increasingly abstract representation of the input data. This design is fundamental to the capability of neural networks to learn complex relationships and make accurate predictions.

Mapping a NN architecture onto an AIE architecture utilizes vector operations to optimize computational efficiency and performance. In our implementation, the forward pass of a layer is executed by leveraging AIE's vector capabilities. Initially, input data and weights are represented as vectors, facilitating the simultaneous processing of multiple elements in parallel. The key operation involves multiplying the input vector by the weight vector, performed through a vectorized multiplication operation. This approach accelerates the computation by allowing the AIE to handle multiple multiplications at once, which is significantly faster than traditional scalar methods. 

\section{Evaluation Results}
\label{sec:results}

\begin{figure*}[h]
  \centering
  \includegraphics[width=\linewidth, trim={0 110 0 0},clip]{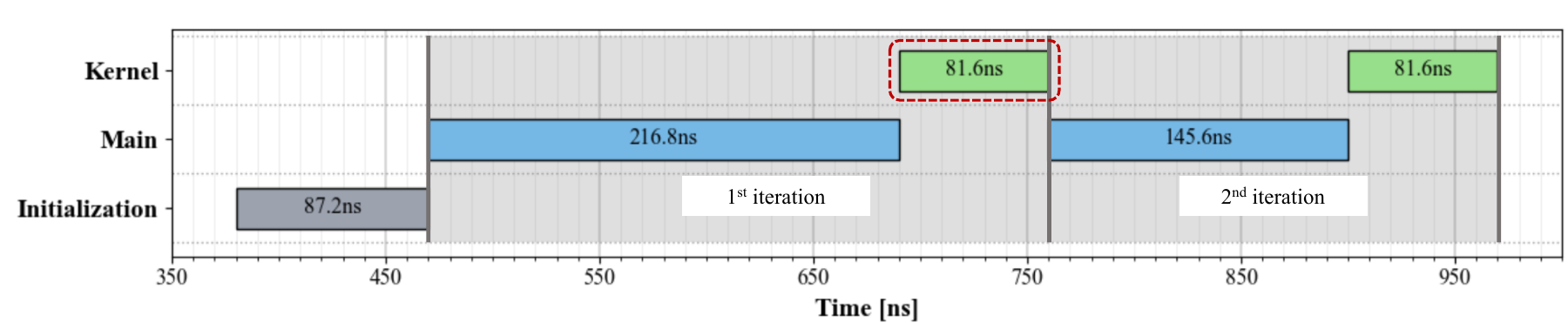}
  \caption{Execution timeline of functions on the AI Engine. The initialization phase is followed by two iterations of the main program, each invoking the neural network kernel. Measured execution times for initialization, main execution, and NN run are shown.}
  \label{fig:perf_results}
\end{figure*}

\begin{figure}[h]
  \centering
  \includegraphics[width=\linewidth, trim={0 0 370 0},clip]{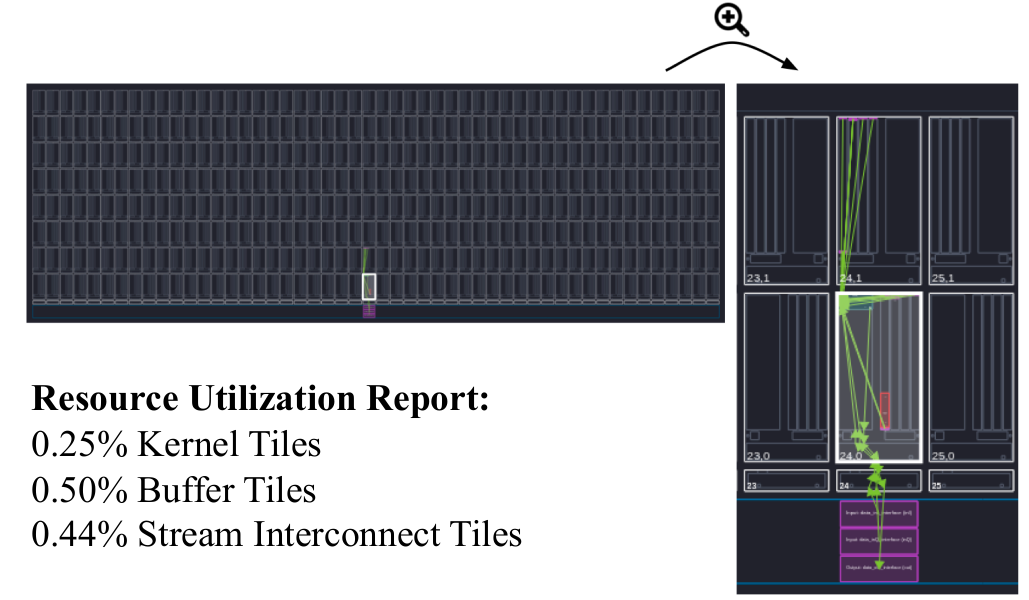}
  \caption{AI Engine resource utilization for the state discrimination neural network kernel mapped onto the AMD VCK190 device. The figure shows the placement of kernel, buffer, and stream interconnect tiles across the AI Engine array. Only a small fraction of available resources is used, highlighting the compact footprint of the proposed design.}
  \label{fig:ru}
\end{figure}


We evaluate our design using AMD Vitis simulation tools for the AMD VCK190 evaluation board. To support efficient data movement, we implement a streaming pipeline in the PL that transfers data between off-chip DDR memory and the AI Engine using AXI Stream interfaces. The pipeline consists of memory-mapped–to-stream (MM2S) and stream-to-memory-mapped (S2MM) modules, enabling continuous data streaming to and from the AI Engine. Our results are obtained with the AIE operating at 1250 MHz. 

\textbf{Latency:} Figure~\ref{fig:perf_results} shows the measured execution latency of the state discrimination application. The application execution is divided into three logical phases. The Initialization phase performs one-time setup operations, such as configuring DMA engines, initializing data structures, and preparing input and output buffers in memory. The Main function acts as the control loop of the application: it manages iteration-level execution, initiates data transfers between DDR memory and the programmable logic, and triggers execution of the AI Engine kernels. The Kernel represents the NN computation mapped onto the AI Engine, where the bulk of arithmetic operations are performed. Unlike the initialization and control logic, the kernel executes on dedicated AI Engine hardware and exhibits consistent execution latency across iterations. 

We emphasize that our performance comparison with related machine-learning–based approaches focuses on the \textbf{kernel execution latency}, as the kernel corresponds directly to the neural network (NN) inference computation. In particular, QubicML~\cite{2024arXiv240618807V} reports an NN inference latency of 54 ns, which is comparable to the 81.6 ns kernel latency measured in our implementation. While our kernel latency is higher, it is important to note that it is achieved within a fully programmable AI Engine pipeline and includes deterministic execution behavior.

The additional overhead observed in the \texttt{Main} function primarily arises from control logic and data orchestration. We expect that this overhead can be largely hidden in a pipelined execution model by overlapping control execution with data streaming, for example, by initiating streaming operations while waiting for kernel results. Such overlap is common in streaming architectures and can significantly reduce the effective end-to-end latency.

A key advantage of our approach is that the \textbf{kernel latency is expected to remain constant} as the dataset size or neural network architecture scales. This behavior follows from our use of vector-level parallelism on the AI Engine, which allows the NN computation to scale with available vector resources rather than input size. As a result, the proposed design is well-suited for larger qubit systems and naturally extends to higher-dimensional quantum states, such as qutrits and qudits, without increasing kernel latency.

We note, however, that additional work is required to fully hide the control and data-movement overheads. In particular, tighter coordination between data streaming, kernel execution, and the surrounding control system components will be necessary to maximize overlap and minimize exposed latency. We leave a detailed exploration of these optimizations to future work.

\textbf{Resource Utilization:} Figure~\ref{fig:ru} illustrates the AIE resource utilization of the proposed state discrimination NN kernel. The implementation occupies \textbf{0.25\% of kernel tiles}, \textbf{0.50\% of buffer tiles}, and \textbf{0.44\% of stream interconnect tiles} on the AMD VCK190 device. The placement view shows that the design is confined to a localized region of the AIE array, leaving the majority of tiles unused.

This low resource footprint demonstrates that the proposed kernel is lightweight and highly scalable. The limited consumption of AIE resources allows additional kernels or larger neural network instances to be instantiated without exceeding device capacity. Moreover, the availability of unused tiles enables future extensions such as deeper models, parallel kernel replication, or multi-channel processing to further improve throughput.

Overall, the resource utilization results indicate that the proposed design efficiently leverages the AIE architecture while preserving substantial headroom for scaling to larger problem sizes or more complex quantum state representations.

\textbf{Power Consumption:} Power consumption is estimated using the AMD Vitis power analysis tool. The results indicate that the proposed design consumes \textbf{0.593 W} when executing the NN kernel on the AIE. This power is dominated by \textbf{dynamic power}, with \textbf{0.092 W} attributed to the AIE core computation and \textbf{0.501 W} to on-chip memory activity. The low overall power consumption is consistent with the compact resource footprint of the design, which utilizes only a single AI Engine core (0.25\% of available cores) and minimal PL or NoC resources.

While power consumption is not the primary optimization target for quantum control and readout pipelines, it remains an important secondary metric. In many quantum computing platforms, classical control hardware operates outside the cryogenic environment, where strict power budgets are less constraining than for qubit hardware itself. In this context, the reported sub-watt power consumption is not a limiting factor for deployment.

However, power efficiency becomes increasingly relevant as quantum systems scale and require higher degrees of parallel classical processing, for example when controlling larger numbers of qubits or performing real-time feedback and decoding. The low power consumption observed in our implementation suggests that the proposed approach can scale without introducing significant thermal or energy overheads on the classical control side.

\section{Conclusions and Future Work}
\label{sec:conc}

In this work, we presented a low-latency neural network–based state discrimination pipeline implemented on the AMD AI Engine. Using a streaming execution model, we demonstrated deterministic kernel execution with a measured latency of 81.6 ns at 1250 MHz, while maintaining a compact resource footprint and sub-watt power consumption. Our results show that the neural network kernel dominates the computational workload and provides stable latency across iterations, making it suitable for real-time quantum computing control and readout applications.

Future work will focus on further reducing end-to-end latency by overlapping control execution and data movement with kernel execution through tighter coordination of the streaming pipeline and control logic. We also plan to scale the design to support larger neural network architectures and higher-dimensional quantum systems, including qutrits and qudits, leveraging the vector parallelism of the AI Engine. In addition, we intend to validate the proposed approach through experiments with real qubit hardware, integrating the AI Engine–based inference pipeline into a complete quantum control stack to assess performance under realistic operating conditions. Finally, extending the design to support multi-kernel parallelism and closed-loop feedback will be explored to enable more advanced real-time quantum control scenarios.

An important direction for future work is to investigate emerging hardware platforms, such as the recently announced AMD Versal RF Series adaptive SoCs~\cite{amd_versal_rf_2023} -- that integrate both high-performance AI acceleration and RF data conversion on a single board. Such integration would eliminate the need to stream data across multiple boards, reducing system complexity, communication latency, and synchronization overhead.

\section{Acknowledgments}
This work was supported by the Laboratory Directed Research and Development Program of Lawrence Berkeley National Laboratory under U.S. Department of Energy Contract No. DE-AC02-05CH11231.



\bibliographystyle{ACM-Reference-Format}
\bibliography{biblio}

@article{preskill2018quantum,
  title={Quantum computing in the NISQ era and beyond},
  author={Preskill, John},
  journal={Quantum},
  volume={2},
  pages={79},
  year={2018},
  publisher={Institute of Physics}
}

@article{arute2019quantum,
  title={Quantum supremacy using a programmable superconducting processor},
  author={Arute, Frank and Arya, K and Babbush, Ryan and Bacon, Derek and Bardin, J and et al.},
  journal={Nature},
  volume={574},
  pages={505--510},
  year={2019},
  publisher={Nature Publishing Group},
  doi={10.1038/s41586-019-1666-5}
}

@article{corcoles2019generating,
  title={Generating entanglement in superconducting qubits via mid-circuit measurements},
  author={C{\'o}rcoles, Adriana D and et al.},
  journal={Quantum Science and Technology},
  volume={4},
  number={2},
  pages={025007},
  year={2019},
  publisher={IOP Publishing},
  doi={10.1088/2058-9565/aaeb72}
}

@article{bourassa2020high,
  title={High-fidelity readout of a superconducting qubit},
  author={Bourassa, M. and et al.},
  journal={Nature},
  volume={580},
  number={7803},
  pages={232--236},
  year={2020},
  publisher={Nature Publishing Group},
  doi={10.1038/s41586-020-2166-1}
}

@article{ibm2020quantum,
  title={Quantum volume: a measure of quantum computer performance},
  author={Kandala, Abhinav and Mezzacappa, Antonio and et al.},
  journal={IBM Journal of Research and Development},
  volume={64},
  number={3/4},
  pages={1--11},
  year={2020},
  publisher={IBM},
  doi={10.1147/JRD.2020.3018740}
}

@article{bausch2024learning,
  title={Learning high-accuracy error decoding for quantum processors},
  author={Bausch, J. and Senior, A. W. and Heras, F. J. H. and et al.},
  journal={Nature},
  volume={635},
  pages={834--840},
  year={2024},
  doi={10.1038/s41586-024-08148-8}
}

@ARTICLE{2019arXiv191205114R,
       author = {{Reilly}, D.~J.},
        title = "{Challenges in Scaling-up the Control Interface of a Quantum Computer}",
      journal = {arXiv e-prints},
         year = 2019,
        month = dec,
          eid = {arXiv:1912.05114},
        pages = {arXiv:1912.05114},
          doi = {10.48550/arXiv.1912.05114},
archivePrefix = {arXiv},
       eprint = {1912.05114},
 primaryClass = {quant-ph},
       adsurl = {https://ui.adsabs.harvard.edu/abs/2019arXiv191205114R}
}

@misc{amd2020ai,
  title={AMD AI Engine: Accelerating Machine Learning and AI Workloads},
  author={AMD, Inc.},
  year={2020},
  url={https://www.amd.com/en/technologies/amd-ai-engine}
}

@ARTICLE{2024arXiv240209532C,
       author = {{Cao}, Shuxiang and {Shao}, Zhen and {Zheng}, Jian-Qing and {Alghadeer}, Mohammed and {Fasciati}, Simone D and {Piscitelli}, Michele and {Spring}, Peter A and {Wang}, Shiyu and {Tamate}, Shuhei and {Vora}, Neel and {Xu}, Yilun and {Huang}, Gang and {Nowrouzi}, Kasra and {Nakamura}, Yasunobu and {Siddiqi}, Irfan and {Leek}, Peter and {Lyons}, Terry and {Bakr}, Mustafa},
        title = "{Superconducting qubit readout enhanced by path signature}",
      journal = {arXiv e-prints},
         year = 2024,
        month = feb,
          eid = {arXiv:2402.09532},
        pages = {arXiv:2402.09532},
          doi = {10.48550/arXiv.2402.09532},
archivePrefix = {arXiv},
       eprint = {2402.09532},
 primaryClass = {quant-ph},
       adsurl = {https://ui.adsabs.harvard.edu/abs/2024arXiv240209532C}
}

@article{goss2024extending,
  title={Extending the computational reach of a superconducting qutrit processor},
  author={Goss, N. and Ferracin, S. and Hashim, A. and et al.},
  journal={npj Quantum Information},
  volume={10},
  pages={101},
  year={2024},
  publisher={Nature Publishing Group},
  doi={10.1038/s41534-024-00892-z}
}

@article{shammah2024open,
  title={Open hardware solutions in quantum technology},
  author={Shammah, Nathan and Roy, Anurag Saha and Almudever, Carmen G. and Bourdeauducq, Sébastien and Butko, Anastasiia and Cancelo, Gustavo and Clark, Susan M. and Heinsoo, Johannes and Henriet, Loïc and Huang, Gang and Jurczak, Christophe and Kotilahti, Janne and Landra, Alessandro and LaRose, Ryan and Mari, Andrea and Nowrouzi, Kasra and Ockeloen-Korppi, Caspar and Prawiroatmodjo, Guen and Siddiqi, Irfan and Zeng, William J.},
  journal={APL Quantum},
  volume={1},
  number={1},
  pages={011501},
  year={2024},
  publisher={AIP Publishing},
  doi={10.1063/5.0180987}
}

@ARTICLE{2024arXiv240618807V,
       author = {{Vora}, Neel R. and {Xu}, Yilun and {Hashim}, Akel and {Fruitwala}, Neelay and {Nguyen}, Ho Nam and {Liao}, Haoran and {Balewski}, Jan and {Rajagopala}, Abhi and {Nowrouzi}, Kasra and {Ji}, Qing and {Whaley}, K. Birgitta and {Siddiqi}, Irfan and {Nguyen}, Phuc and {Huang}, Gang},
        title = "{ML-Powered FPGA-based Real-Time Quantum State Discrimination Enabling Mid-circuit Measurements}",
      journal = {arXiv e-prints},
         year = 2024,
        month = jun,
          eid = {arXiv:2406.18807},
        pages = {arXiv:2406.18807},
          doi = {10.48550/arXiv.2406.18807},
archivePrefix = {arXiv},
       eprint = {2406.18807},
 primaryClass = {quant-ph},
       adsurl = {https://ui.adsabs.harvard.edu/abs/2024arXiv240618807V}
}

@article{PhysRevApplied.10.034040,
  title = {Rapid High-fidelity Multiplexed Readout of Superconducting Qubits},
  author = {Heinsoo, Johannes and Andersen, Christian Kraglund and Remm, Ants and Krinner, Sebastian and Walter, Theodore and Salath\'e, Yves and Gasparinetti, Simone and Besse, Jean-Claude and Poto\ifmmode \check{c}\else \v{c}\fi{}nik, Anton and Wallraff, Andreas and Eichler, Christopher},
  journal = {Phys. Rev. Appl.},
  volume = {10},
  issue = {3},
  pages = {034040},
  numpages = {14},
  year = {2018},
  month = {Sep},
  publisher = {American Physical Society},
  doi = {10.1103/PhysRevApplied.10.034040},
  url = {https://link.aps.org/doi/10.1103/PhysRevApplied.10.034040}
}

@book{helstrom1976quantum,
  title={Quantum Detection and Estimation Theory},
  author={Helstrom, C.W.},
  isbn={9780123400505},
  lccn={75026347},
  series={Mathematics in Science and Engineering : a series of monographs and textbooks},
  url={https://books.google.com/books?id=1N5QrSnsEgcC},
  year={1976},
  publisher={Academic Press}
}

@article{magesan2015machine,
  title={Machine learning for discriminating quantum measurement trajectories and improving readout},
  author={Magesan, Easwar and Gambetta, Jay M and C{\'o}rcoles, AD and Steffen, Matthias},
  journal={Physical Review Letters},
  volume={114},
  number={20},
  pages={200501},
  year={2015},
  publisher={APS}
}

@article{ghojogh2019linear,
  title={Linear and Quadratic Discriminant Analysis: Tutorial},
  author={Ghojogh, Benyamin and Crowley, Mark},
  journal={arXiv preprint arXiv:1906.02590},
  year={2019}
}

@article{lienhard2022deep,
  title={Deep-neural-network discrimination of multiplexed superconducting qubit states},
  author={Lienhard, Benjamin and Antwi, Sarah and Roberts, David and others},
  journal={Physical Review Applied},
  volume={17},
  number={1},
  pages={014024},
  year={2022},
  publisher={APS}
}

@ARTICLE{2022PhRvP..17a4024L,
       author = {{Lienhard}, Benjamin and {Veps{\"a}l{\"a}inen}, Antti and {Govia}, Luke C.~G. and {Hoffer}, Cole R. and {Qiu}, Jack Y. and {Rist{\`e}}, Diego and {Ware}, Matthew and {Kim}, David and {Winik}, Roni and {Melville}, Alexander and {Niedzielski}, Bethany and {Yoder}, Jonilyn and {Ribeill}, Guilhem J. and {Ohki}, Thomas A. and {Krovi}, Hari K. and {Orlando}, Terry P. and {Gustavsson}, Simon and {Oliver}, William D.},
        title = "{Deep-Neural-Network Discrimination of Multiplexed Superconducting-Qubit States}",
      journal = {Physical Review Applied},
         year = 2022,
        month = jan,
       volume = {17},
       number = {1},
          eid = {014024},
        pages = {014024},
          doi = {10.1103/PhysRevApplied.17.014024},
archivePrefix = {arXiv},
       eprint = {2102.12481},
 primaryClass = {quant-ph},
       adsurl = {https://ui.adsabs.harvard.edu/abs/2022PhRvP..17a4024L}
}

@article{PRXQuantum.5.030339,
  title = {Efficient Long-Range Entanglement Using Dynamic Circuits},
  author = {B\"aumer, Elisa and Tripathi, Vinay and Wang, Derek S. and Rall, Patrick and Chen, Edward H. and Majumder, Swarnadeep and Seif, Alireza and Minev, Zlatko K.},
  journal = {PRX Quantum},
  volume = {5},
  issue = {3},
  pages = {030339},
  numpages = {20},
  year = {2024},
  month = {Aug},
  publisher = {American Physical Society},
  doi = {10.1103/PRXQuantum.5.030339},
  url = {https://link.aps.org/doi/10.1103/PRXQuantum.5.030339}
}

@misc{xilinx2023versal,
  title = {Xilinx Versal},
  author = {{Xilinx}},
  year = {2023},
  url = {https://japan.xilinx.com/products/silicon-devices/acap/versal.html},
  note = {Accessed: 2023-10-10}
}

@misc{xilinx2023vck190,
  title = {VCK190 Evaluation Kit},
  author = {{Xilinx}},
  year = {2023},
  url = {https://www.xilinx.com/products/boards-and-kits/VCK190.html},
  note = {Accessed: 2023-10-10}
}

@misc{amd_versal_rf_2023,
  author       = {{Advanced Micro Devices, Inc.}},
  title        = {AMD Introduces Versal RF Series Adaptive SoCs Offering the Industry’s Highest Compute in a Single-Chip Device with Integrated Direct RF Sampling Converters},
  year         = {2023},
  howpublished = {\url{https://ir.amd.com/news-events/press-releases/detail/1231/amd-introduces-versal-rf-series-adaptive-socs-offering-the-industrys-highest-compute-in-a-single-chip-device-with-integrated-direct-rf-sampling-converters}},
  note         = {Accessed: 2026-01-12}
}

\end{document}